\documentclass[conference]{IEEEtran}

\usepackage{cite}
\usepackage{amsmath,amssymb,amsfonts}
\usepackage{algorithmic}
\usepackage{array}
\usepackage{textcomp}
\usepackage{xcolor}

\usepackage{graphicx}
\usepackage{grffile}

\usepackage{tabularx}

\usepackage{booktabs} 

\usepackage[inline]{enumitem}

\usepackage[newfloat,frozencache=true,cachedir=.]{minted} 
\setminted{
    linenos=false,
    fontsize=\footnotesize,
    autogobble,
}

\usepackage{subcaption}
\newenvironment{code}{\captionsetup{type=listing}}{}
\SetupFloatingEnvironment{listing}{
name=Code,
placement=ht,
}

\usepackage{menukeys}

\newcommand{\frisbee}[1]{{\em Frisbee}}

\def\BibTeX{{\rm B\kern-.05em{\sc i\kern-.025em b}\kern-.08em
    T\kern-.1667em\lower.7ex\hbox{E}\kern-.125emX}}

\begin{document}

\title{Event-Driven Testing For Edge Applications}

\author
{\IEEEauthorblockN{Fotis Nikolaidis}
\IEEEauthorblockA{Institute of Computer Science, \\ FORTH (ICS)\\
Heraklion, Greece\\
fnikol@ics.forth.gr }
\and
\IEEEauthorblockN{Antony Chazapis}
\IEEEauthorblockA{Institute of Computer Science, \\ FORTH (ICS)\\
Heraklion, Greece\\
chazapis@ics.forth.gr}
\and
\IEEEauthorblockN{Manolis Marazakis}
\IEEEauthorblockA{Institute of Computer Science, \\ FORTH (ICS)\\
Heraklion, Greece\\
maraz@ics.forth.gr}
\and
\IEEEauthorblockN{Angelos Bilas}
\IEEEauthorblockA{Institute of Computer Science, \\ FORTH (ICS)\\
Heraklion, Greece\\
bilas@ics.forth.gr}
}

\maketitle

\begin{abstract}
With the rise of the Internet of Things (IoT) and Edge computing, a considerable amount of system services are moving from reliable Cloud data centers to less reliable infrastructures closer to the end-users. However, the constrained resources, unreliable communication, and varying operating conditions of IoT pose significant complexities for software vendors in testing their applications. Although several emulators exist for testing IoT systems, numerous issues can be pointed out, such as lacking support for logical dependencies and advanced fault injection capabilities, requiring manual validation of the system's behavior, or focusing on a specific platform and language. To address these limitations, we propose Frisbee: a framework for the automated testing of IoT applications. Frisbee accelerates the testing process by simplifying the spin-up of distributed virtual testbeds over Kubernetes and the description of `what-if' scenarios that integrate complex sequences of workloads and faultloads. 
\end{abstract}


\section{Introduction}

The Internet-of-Things (IoT) poses one of the most significant and widespread systems, where errors directly impact people’s lives. Testing and validating is how one deals with errors; but testing and validating a planetary-scale, heterogeneous, and ever-growing ecosystem has its own challenges. Although individual IoT devices have limited functionalities and are easy to test, the testing can become remarkably complex when we consider thousands of devices deployed in dynamic topologies, unreliable connectivity, and device or protocol heterogeneity.

For this reason, a considerable amount of work has been done on virtual testbeds that replicate the physical properties of an IoT infrastructure. Virtual testbeds like Fogify~\cite{symeonides2020fogify} and IOTier~\cite{nikolaidis2021iotier} can transform a cluster of lab machines into a Virtual Environment that supports complex topologies comprised of heterogeneous resources and versatile network capabilities with multiple QoS criteria. However, the testbeds only provide the testing infrastructure. Testers will still have to think long and hard about how to evaluate a complex distributed system such as an IoT deployment and how they want to structure their test operations. 

The first challenge for testers is to bring the System Under Testing (SUT) into an initial working state. This can be a particularly involved process since IoT consists of multiple dependent services that must be boot in a specific order. Unfortunately, almost all developed IoT emulators are time-driven, meaning that all actions are scheduled in fixed intervals without accounting for the system’s state. As a result, even slight changes in the initial conditions across executions (e.g., CPU load) can lead to premature or delayed actions that hinder the credibility of the test case. 

The second challenge is to assess how the system reacts to different workload and faultload configurations. Despite the abundance of workload-generating benchmarks and fault-injecting Chaos tools, the lack of a unified tool has led to severe fragmentation in testing methodology and poor comparability of results. Prior studies approached the subject by submitting a running system into randomized, negative behavior, like shutting down nodes, introducing CPU and network stressors, or injecting faults. The goal is to uncover shallow (independent) bugs~\cite{kao1996define} early in the development cycle~\cite{abedi2017conducting, leitner2016patterns}. However, random strategies are unlikely to reveal “deep” failures which manifest under specific operating conditions~\cite{giuffrida2013edfi}, e.g., quorum bugs that appear when connectivity is broken during the synchronization of a replica-set~\cite{mongobug}. Not surprisingly, prior studies have reported fault injection campaigns with no faults activated in as many as 40\% of the experiments~\cite{ng2001design, simao2010fault}. This untested code, often the error recovery code, will tend to be an error-prone part of a system~\cite{10.1145/3447851.3458738}. 

Finally, neither existing Chaos tools nor Virtual Testbeds provide the means for assessing the impact of a fault. Consequently, the tester has to design and maintain monitoring mechanisms to ensure that the test case completes without failures, without correctness issues~\cite{jepsen}, and without SLO violations~\cite{10.1145/1807128.1807153}. We argue that these shortcomings tend to make the end-to-end testing come late in the development cycle; when bugs are revealed, reworked, and the additional retesting cost is high and impacts the product release~\cite{10.1007/978-3-030-58858-8_26, miller2021four}. 

This paper presents \frisbee{}, the first unified platform to support end-to-end testing over Kubernetes. Unlike all prior tools, \frisbee{} delivers a fully automated testing environment that can spin up multiple containers on-demand, awaiting containers to start up and load data (plus dependent containers), then manipulating containers (e.g., via fault injection), and validating that the system behaves as expected. To achieve that, \frisbee{} integrated a combination of Cloud platforms, event-driven workflows, Chaos toolkits, and a well-integrated distributed monitoring stack that provides insights and alerting about the system’s state.

Our contributions are:
\begin{itemize}
\item A DSL that combines complex workloads with faultloads.

\item A methodology for execution-driven fault injection that allows injecting a controlled and predetermined faultload distribution as the system executes at runtime.

\item A design for repurposing existing monitoring tools for the automated assertion of a distributed system’s behavior and side effects.

\item A set of patterns for Load testing, SSTable Corruption, and Goroutine Leaking, for a Cloud database that supports IoT applications.
\end{itemize}


\section{The Frisbee Framework}

\subsection{Features}
With \frisbee{}, developers of IoT applications, who seek to explore application performance and reliability under various operational conditions, can overcome the challenges mentioned above by being provided with the following features:

\noindent
\textbf{Extensibile and Portable Testbed}
We carefully designed \frisbee{} to be easily extensible, portable, and practical. We implemented the \frisbee{} core features to operate in Kubernetes to decouple our framework from any system-specific constraints and provide similar environments for dev, test, and production. Doing so, \frisbee{} scenarios are reproducible across local workstations (e.g., minikube, microk8s),  bare-metal cluster setups (e.g., k8s), as well as in the Cloud deployments (e.g., EKS). 

\noindent
\textbf{Event-Driven Initialization, Workload and Faultload}
For controllability and precision of the executed actions, \frisbee{} provides direct access to the runtime environment, allowing users to build execution-driven scenarios. This way, testers can i) spin up the dependency stack of the SUT in a deterministic manner; ii) build complex workloads with dynamically changing request patterns; and iii) surgically inject faults into specific locations without introducing spurious faults that may compromise the validity of the results.

\noindent
\textbf{Failure-Aware Semantics} 
Another advantage of managing the SUT and the failure injection from within the same workflow is the ability to distinguish an unexpected service failure (i.e., due to a bug in the application's code) from an expected service failure (i.e., part of a Chaos experiment). That is a highly-required feature in resiliency testing~\cite{tolosana2010adaptive} since in the first case, the test should abort immediately, whereas in the second case, the fault does not constitute a culpable failure, and the test should continue. 

\noindent
\textbf{Contextualized Visualizations:}  Particular emphasis is given to supplying additional interpretation and contextual data that help testers visually correlate the observed behavior with a root event. For instance, an annotation marking a network fault's injection could help explain an otherwise enigmatic drop in the system's throughput. To achieve that, when \frisbee{} performs an action (e.g., create a service, inject a fault), it interacts with Grafana to annotate the action's beginning and ending.

\noindent
\textbf{Hierarchical Assertions}
Since testing workflows are defined hierarchically (e.g., clusters managing hundreds of services), it is rather natural to consider the partitioning of large verification problems into a set of properties associated locally with the entities of the workflow hierarchy~\cite{kasuya2007verification}. Instead of large and complex assertions, \frisbee{} promotes the decomposition of complex properties into simpler assertions, provable at the action level. 

\subsection{System Overview}

\begin{figure}[htbp]
    \centering
    \includegraphics[page=1, width=\columnwidth]{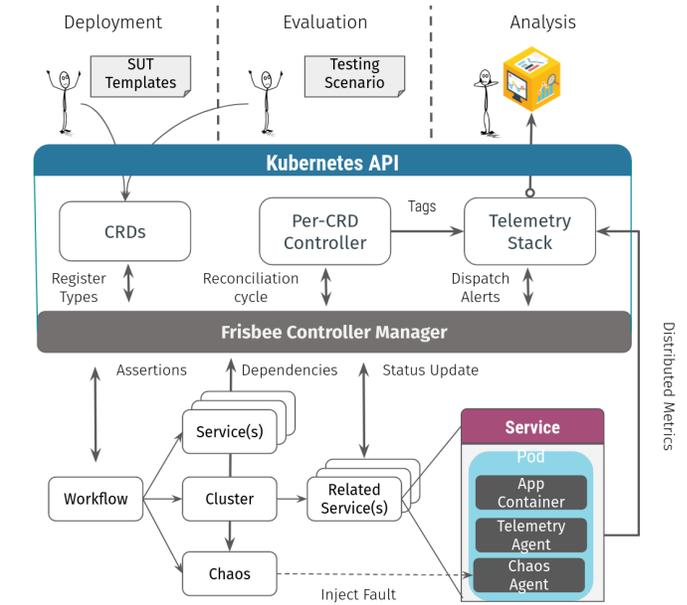}
    \caption{Frisbee architecture. Given a template describing the system under test and a workflow describing the experiment, \frisbee{} interfaces with Kubernetes to run the experiment. Notice the loop among controllers, telemetry stack, and kubernetes API. }
    \label{fig:architecture:overview}
\end{figure}

    Figure~\ref{fig:architecture:overview} depicts a high-level overview of the \frisbee{} architecture. Built over the Operator pattern of Kubernetes~\cite{operator}, \frisbee{} consists of Custom Resource Definitions (CRDs) and a collection of controllers. A CRD defines a new, unique object Kind in the Kubernetes API, listing out all of the configurations for describing the desired state of that object. A controller is a software loop that watches the Kubernetes API for changes and, if necessary, takes actions to reconcile the expressed desired state and the current state of an object. 
    
    The experiment starts with specifying the testing resources in a YAML file. The most important are the templates that describe the System Under Testing (SUT) and the action workflow that drives the testing process. The workflow is a directed acyclic graph (DAG) whose nodes represent jobs that change the execution environment and can be either constructive (e.g., create a new service) or destructive (e.g., kill a service, inject a network fault). The edges represent a bidirectional parent-child relationship between jobs. Every parent is responsible for overseeing the creation and the management of its constituent jobs and updating its status accordingly. This hierarchy facilitates decentralized testing (fewer jobs to be managed by the central workflow controller) and provides context-based execution for decoupling the outcome of the test case from individually failing services.
    
    Users do not directly interact with the Frisbee Controller Manager. Instead, when the scenario is ready for deployment, the user submits it to the Kubernetes API via the standard kubectl tool. In turn, the Kubernetes API submits the scenario to the registered \frisbee{} controller, which checks for malformed dependencies and errors in expressions (e.g., assertions, conditional branches) before the actual deployment. This validation enables us to avoid loops and guarantee that a test case should not continue running forever, preventing other test cases from execution. Subsequently, Kubernetes triggers a reconciliation cycle to the workflow controller. During this cycle, the controller traverses the workflow graph and executes the listed jobs. In general, jobs are executed immediately unless the tester has defined scheduling constraints or logical dependencies to other jobs. In this case, the execution is postponed until one of the events of Table~\ref{tab:events} is fired. 
    
    Upon a fired event, the controller starts a new reconciliation cycle during which it interacts with the Kubernetes API (e.g., create, update or delete objects) or performs some action on other systems, e.g., disseminate policies to \frisbee{} agents at the execution layer. \frisbee{} uses sidecar agents to interact with application containers~\cite{burns2018designing,196346}, in a non-intrusive manner. Agents exist for CPU or network throttling, monitoring system resources (CPU, memory, network, IOPS), monitoring application metrics (e.g., quorum state, replication role), and injecting faults (even network partition are rules set by a network agent to the routing table of the pod).
    
\begin{table}[htpb]
\centering
\begin{tabularx}{\columnwidth}{|c||X|}
{\textbf{Event}}  &{\textbf{Description}} \\
\hline    
Time-driven     & Fired after an elapsed time measured by the controller.  \\

State-driven     & Fired by the Kubernetes API when managed objects have state changes, errors, or other messages that should be broadcast to the system.    \\

Metrics-driven     & Fired by the telemetry stack when the outcome of statistical analysis on the collected metrics matches a given rule.  \\

Tag-events     & Fired when one controller passes contextual information to another controller.   
\end{tabularx}
\caption{Events used to drive assertions, conditional loops, and other places involving execution-driven knowledge.}
\label{tab:events}
\end{table}


\section{Implementation Details}
    \subsection{Scenario Modeling}
    A Frisbee scenario describes a set of dependent actions that collectively describe the testing process. The scenario defines three important properties:
    \begin{enumerate*}
        \item a list of actions that drive the testing process,
        \item the preconditions of the runtime before each action, and
        \item the desired state of the runtime after each action.
    \end{enumerate*}

    \textbf{Actions} covers a full range of operations required to test a distributed system, and consists of five comprehensive and fine-grained abstractions: services, clusters, faults, calls, and checkpoints. Descriptions are in Table~\ref{tab:actions}. Under the hood, actions invoke templates that provide solid definitions of the SUT. This strategy allows us to create a library of frequently-used specifications and use them to generate objects on-demand throughout the experiment. When called without parameters, templates generate services initialized with reasonable defaults. With parameters, templates generate the customized configuration by replacing the placeholders (denoted as {{...}}) with the given input. Due to the limited space, we only provide the definition of a scenario~\ref{code:scenario}, not the underlying templates.

\begin{table}[htpb]
\centering
\begin{tabularx}{\columnwidth}{|c||X|}
\toprule
{\textbf{Action}}  &{\textbf{Description}} \\
\hline    
Service     & Standard representation of a containerized application.  \\

Cluster     & Abstraction for managing multiple services in a shared execution context.    \\

Fault     & Condition that disturbs a running service, network, or storage.  \\

Call     & Command executed by the controller to a running service.  \\

Checkpoint  &  Store the status and the metrics of the SUT at the given time. 
\end{tabularx}
\caption{Using just 5 basic primitives, \frisbee{} covers a full range of operations required to test a distributed system.}
\label{tab:actions}
\end{table}

\begin{code}
\begin{minted}{yaml}
spec:
# Step 0. Provision 4 individual servers
- action: Cluster
  name: masters
  cluster:
    templateRef: cockroach.cluster.master
    instances: 4
# Step 1. Create a cockroach cluster
- action: Call
  name: boot
  depends: { running: [ masters ] }
  call:
    callable: boot
    services: [ .cluster.masters.all ]
# Step 2. Import TPC-C data from node-0
- action: Call
  name: import-workload
  depends: { success: [ boot ] }
  call:
    callable: import-tpcc
    services: [ masters-0 ]          
# Step 3. Wait for 3x replication
- action: Call
  name: wait-for-3x-replication
  depends: { success: [ import-workload ] }
  call:
    callable: wait-for-3x-replication
    services: [ .cluster.masters.all ]
# Step 4. Run TPC-C workload from the node 3.
- action: Call
  name: run-workload
  depends: { success: [ wait-for-3x-replication ] }
  call:
    callable: run-workload
    services: [ masters-3 ]          
# Step5. Partition node 1 from the rest of the nodes;
- action: Chaos
  name: partition0
  depends: { success: [ wait-for-3x-replication ]}
  chaos:
    templateRef: system.chaos.network.partition.partial
    inputs:
      - {   source: masters-0, 
            direction: "to", duration: 10m, 
            dst: "masters-1, masters-2, masters-3" }          
\end{minted}
\captionof{listing}{Scenario demonstrating action with logical dependencies (depends), parameterization (inputs), addressing macros (.cluster), and injection of templates faults.}
\label{code:scenario}
\end{code}

    \subsection{Extensibility and Integration with Chaos Controllers}
    Every action in \frisbee{} is represented by a different CRD and is managed by a separate controller, e.g, the \textit{cluster} action will handled by the \textit{Cluster controller}. Practically, this architecture builds a tree of domain-specific controllers with a top-level controller (the scenario) that dispatch requests based on the type of action. Controllers interact by submitting and receiving requests recorded in the Kubernetes API. In the previous example, the \textit{Scenario controller} will create a \textit{Cluster resource} to the Kubernetes API, that will be subsequently captured by the \textit{Cluster controller}. After handling the request, the \textit{Cluster controller} will modify the \textit{Cluster resource} with the updated status, and the Kubernetes API will notify the \textit{Scenario controller} about the change.
    
    \subsection{Manage Dependencies Between Actions}
    To manage the logical dependencies between actions, the \textit{scenario} controller should inspect the status of the various actions (stored in the Kubernetes API) and, once it comes to the desired state, trigger the following action. This, however, requires the \textit{Scenario} controller to know how to parse the various statuses and understand their semantics. One solution is to let the user specify parsing rules for each action and explicitly model the reaction to every possible outcome. Instead of this laborious task, we have set a common phase-field so that all objects managed by \frisbee{} can consistently communicate where they are in their lifecycle. The values and their meaning are tightly guarded, as shown in Table~\ref{tab:lifecycle}. 

\begin{table}[htpb]
\centering
\begin{tabularx}{\columnwidth}{|c||X|}
\toprule
{\textbf{Phase}}  &{\textbf{Description}} \\
\hline    
Uninitialized     & the request has been accepted by the Kubernetes API, but it is not yet dispatched to the \frisbee{} controller.  \\

Pending     & the request has been accepted by a \frisbee{} controller, but at least one of the constituent jobs has not been created.   \\

Running     & all constituent jobs of the request have been created, and at least one job has not been completed yet.  \\

Success     & all the constituent jobs of a request have been successfully completed.  \\

Failed     & at least one job of the request has terminated in a failure.   
\end{tabularx}
\caption{Lifecycle Semantics. Logical Dependencies are built upon them.}
\label{tab:lifecycle}
\end{table}


    \subsection{Expected versus Unexpected Failures}
\begin{figure}
    \centering
    \includegraphics[width=\columnwidth]{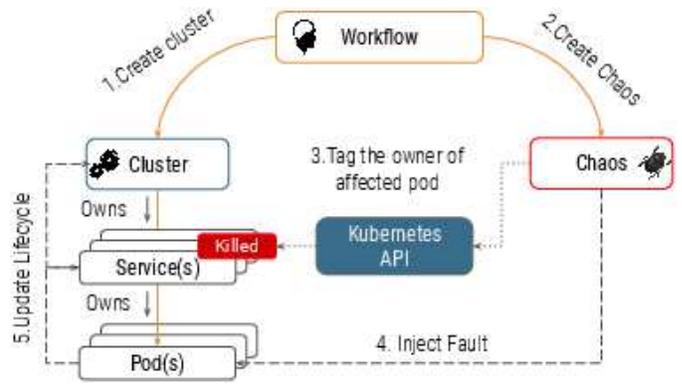}
    \caption{Information propagation for distinguishing unexpected failures caused by application bugs, from expected failures caused by fault-injection jobs. Chaos controller marks the service owning the targeted pod with an indicative tag before it applies the chaos action.}
    \label{fig:ownership}
\end{figure}

    \frisbee{} features a holistic approach to failure detection, propagation, and recovery~\cite{herbein2019mcem}, which provide testers the ability to distinguish between expected and unexpected failure, as shown in Figure~\ref{fig:ownership}. The basis of our mechanism is that almost all fault-injections in Kubernetes happen at a Pod level. Whenever the workflow controller encounters a Chaos action, it traverses the ownership semantics for finding the service responsible for the affected pod and places a \textit{metadata.Chaos} tag on it. This tag describes the type of impending fault (e.g., partition, kill). It then applies the action. If the pod crashes, the failure is propagated to the service via the status updating mechanisms, making the service fail as well. When this happens, the service owner (i.e., cluster) can conclude whether the failure is expected or not by inspecting the service's metadata for the presence of a chaos tag. If there is no tag, the failure is unexpected, and the cluster fails immediately to cut losses and quickly try something else. If there is a tag, the failure is expected, and the behavior is tunable according to the clusters' semantics, e.g., the number of tolerated failures.

    \subsection{Extract State and Performance Metrics}
	Expressions in \frisbee{} can be state-driven or metrics-driven.
    
    \begin{itemize}
        \item \textbf{State Expressions} are used to describe dependencies on the lifecycle of a Kubernetes object (i.e when a service has failed). They consume state events fired by the Kubernetes API. For simplicity, we provide a set of aggregation functions that are callable from within the expressions, e.g., `.state.failed() > 4`. Expressions can also be combined into complex assertions via logical/arithmetic/string comparators. 
        
        \item \textbf{Metrics Expressions} are used to describe dependencies on the performance of a running service (i.e when percentile latency exceeds a given limit). They involve an initial step for setting an alert to Grafana and then checking if the alert is fired. For interoperability, we have adopted the Grafana syntax for writing alerts. For example, "MAX() QUERY(metric, 1m, now) IS ABOVE(70000)" roughly translates as `raise an alert if the max value of the given metric for the period between not and 1 minute ago has been above 70000`. 
    \end{itemize}
    
    \noindent
    The expressions are well-integrated into the \frisbee{} language and are reusable across assertion, conditional loops, and other places involving execution-driven knowledge. Additionally, the extracted output of the expressions can be stored using the \textit{Checkpoint} action. As demonstrated in the Evaluation Section, the \textit{Checkpoint} allows us to take snapshots and use this information in assertions to compare performance metrics before and after an action (i.e., network partition).

    \subsection{Scoped Assertions}
    Every resource managed by \frisbee{} (e.g., services, faults, clusters, cascade) has its own assertions. The main benefit is that we can isolate failed services while allowing the workflow components to synchronize their knowledge and correlate individual failures to top-level events~\cite{gupta2009cifts,di2017logaider}. For state-based expressions, the assertions only consider jobs in the local scope -- those created by the entity. This way, we prevent cross-references on jobs belonging to a different resource,  thus saving the user from never-ending loops. Nonetheless, we cannot wholly exclude cross-references, given that metrics-driven expressions have access to the common telemetry stack. Therefore, we prevent cross-references to a job but not to a job's performance metrics. Regardless of this peculiar case, we still permit top-level assertions to use available assertions at lower hierarchy levels.



\section{Evaluation}
We evaluate \frisbee{} using the CockroachDB. The decision was based on two factors. Firstly, CockroachDB is getting traction from IoT community as it supports numerous required features, such as stability, scalability, SQL-compliance, ACID Transactions, Kubernetes compability, and data locality~\cite{movatic}. Secondly, CockroachDB have open-sourced their own tool (roachtest) for performing large-scale (multi-machine) automated tests, and have a large collection of testing scenarios that we replicate in \frisbee{}.

For the evaluation, we used a Kubernetes cluster of 4 server-grade nodes. Each node has 24-cores (2 Intel Xeon E5-2630v3 CPUs), 128 GB of DDR3 ECC main memory, and 250 GB of locally attached NVMe storage. The machines are connected via 1Gbps links to the top-of-rack switch. We configured CockroachDB and YCSB benchmarks using the default settings. For the monitoring we used the bitnami containers Telegraf, Prometheus, and Grafana. For failure injection, we use Chaos-Mesh~\cite{chaomesh}. 

\textbf{Load Testing} To assess the precision of \frisbee{}’s event-driven strategy, we performed  load testing experiments using a sequence of YCSB workloads. Roughly, the scenario performs the following steps, \begin{enumerate*}
    \item Create a cockroach cluster;
    \item Preload the server with keys;
    \item Run YCSB workload A;
    \item Run YCSB workload B;
    \item Run YCSB workload C;
    \item Run YCSB workload F;
    \item Run YCSB workload D;
    \item Reload the database;
    \item Run YCSB workload E.
\end{enumerate*}

\begin{figure}[ht]
\begin{subfigure}{.5\textwidth}
  \centering
    \includegraphics[page=1, width=\columnwidth]{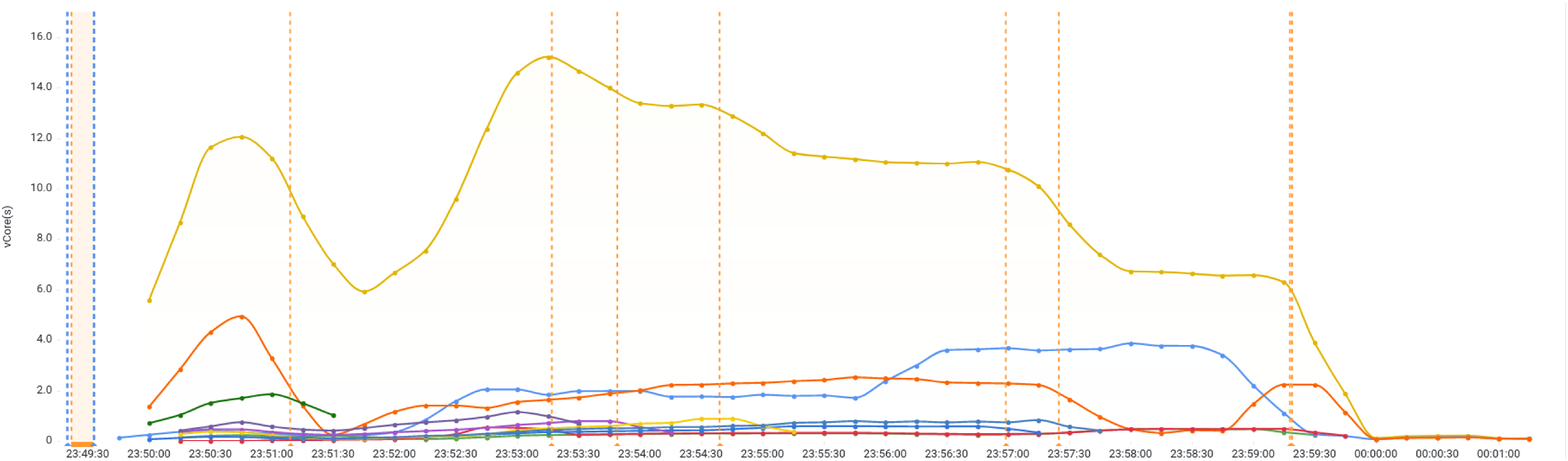}
    \caption{Kubernetes Scheduling, 500 keys}
    \label{fig:baseline:vanilla}
\end{subfigure}

\begin{subfigure}{.5\textwidth}
  \centering
    \includegraphics[page=1, width=\columnwidth]{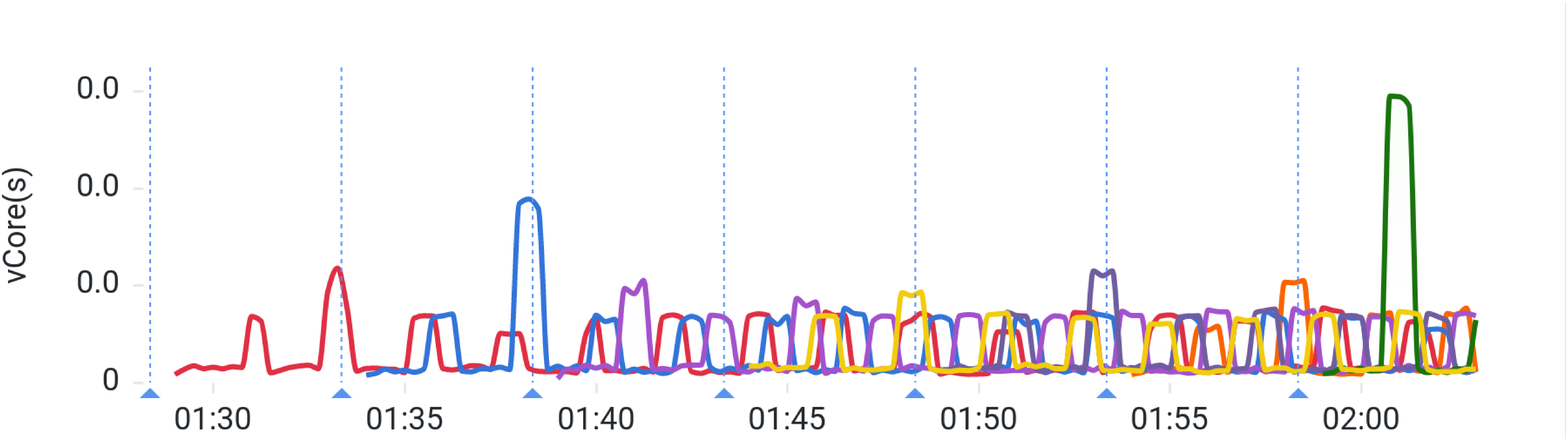}
    \caption{Time-Driven Scheduling, 500 Keys}
    \label{fig:baseline:time-driven_500k}
\end{subfigure}

\begin{subfigure}{.5\textwidth}
  \centering
    \includegraphics[page=1, width=\columnwidth]{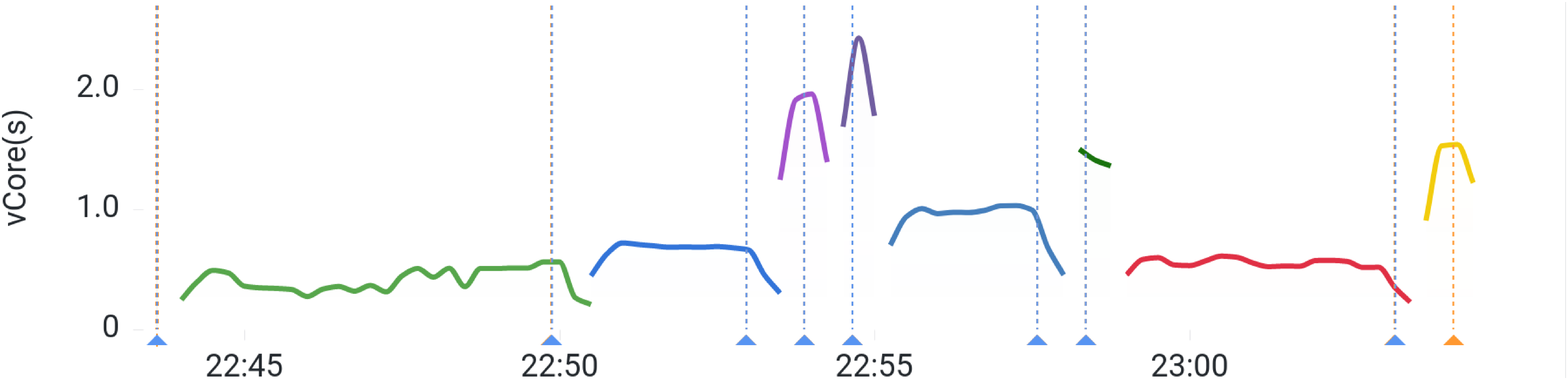}
    \caption{State-Driven Scheduling, 500 Keys}
    \label{fig:baseline:state-driven_500k}
\end{subfigure}

\begin{subfigure}{.5\textwidth}
  \centering
    \includegraphics[page=1, width=\columnwidth]{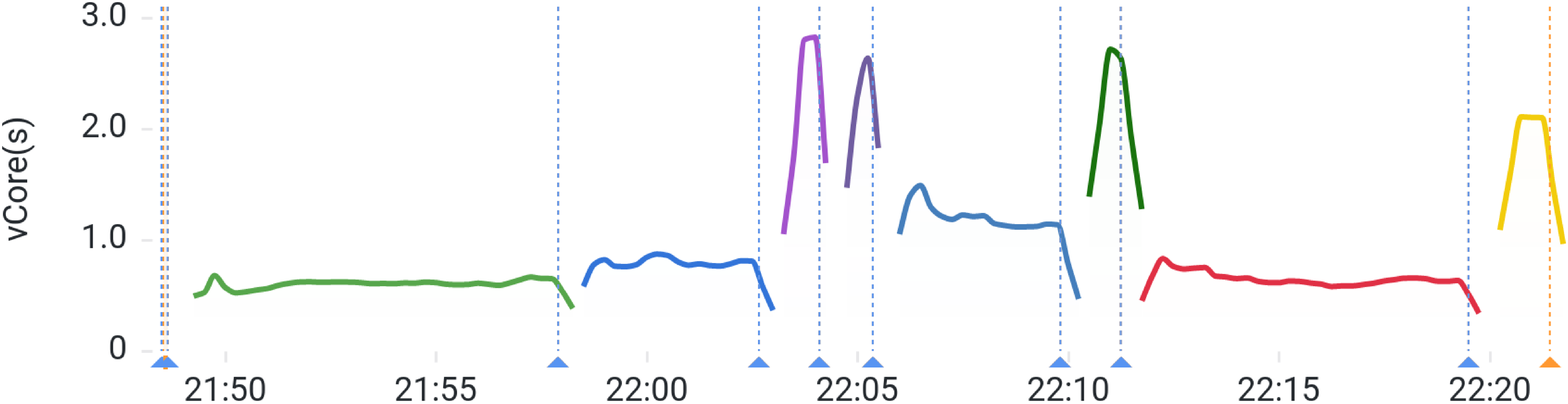}
    \caption{State-Driven Scheduling, 1M keys}
    \label{fig:baseline:state-driven_1m}
\end{subfigure}


\caption{Scheduling Policies for the standard sequence of YCSB workloads (A,B,C,F,D,E)}
\label{fig:scheduling-policies}
\end{figure}

In each experiment, we use a different invocation strategy (Kubernetes Vanilla, Time-Driven, State-Driven), and collect statistics on system utilization (CPU cycles) and application-specific metrics (Go routines) consumed by the database servers. As shown in Figure~\ref{fig:scheduling-policies}, the vanilla Kubernetes policy is to start all jobs in parallel, which makes the experiment extremely flaky. This is because Kubernetes by default always expects the `final', or the desired, state of a resource, thus does not provide primitives for logical dependencies. With the sleeping instrumentation, we see that there are no scheduling guarantees. If the sleeping time is small, subsequent stages of the workflow will overlap, leading to erroneous results. If the sleeping time is large, there is a huge portion of idle time, that is translated to long-running experiments with increased time and energy costs. In contrast, state-driven execution appears to respect the ordering of the jobs, irrespectively of the number of ingested keys.

\textbf{SSTable Corruption}
This scenario is adopted from the roachtest suite~\cite{roach-test-ssstable-corruption} and is designed to test if the storage layer (pebble) can detect corrupted SST files and panic, i.e., fail fast. Roughly, it performs the following steps,
\begin{enumerate*}
    \item provision 3-node cluster
    \item import TPC-C data from the workload node (node 1)
    \item find 6 random SST files on each node (abort if node has fewer)
    \item corrupt each of the 6 SST files on each node
    \item start each node, verify each node panics
    \item if nodes don't panic, run TPC-C workload for up-to 10 mins (node1)
    \item verify cluster panic
    \item fail the test if DeadlineExceeded or Cancelled (workload finished but no panic occurred)
\end{enumerate*}

\begin{figure}[htbp]
    \centering
    \includegraphics[page=1, width=\columnwidth]{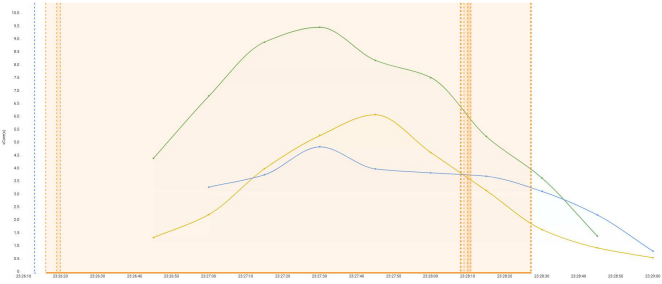}
    \caption{SSTable corruption, with assertions}
    \label{fig:baseline:bitrot}
\end{figure}

As can be seen in Figure~\ref{fig:baseline:bitrot}, \frisbee{} can successfully detect the panicked service and aborts the experiment. It worth mentioning that since this experiment is implemented using the \textit{Call} primitive. In contrast to \textit{Services} which are intended to be `long-running' and are marked using point annotations, \textit{Calls} are intended to be `short-living', and therefore are annotated as regions, marking the beginning and the ending of a \textit{Call}.

\textbf{Goroutine Leak}
This scenario is adopted from the roachtest suite~\cite{roach-test-goroutines-partition} and is designed to test for leaking goroutines after recovering from a network partition. Roughly, it performs the following steps:
\begin{enumerate*}
    \item provision 4-node cluster
    \item import TPC-C data (node 1)
    \item wait for 3x replication
    \item run TPC-C workload from the workload node (node 4)
    \item record maximum number of goroutines observed thus far, i.e., maxSeen
    \item partition node 1 from the rest of the nodes; node 1 can reach other nodes, but no other nodes can reach node 1
    \item continue running with the simulated network partition for 1 hour, regularly updating maxSeen
    \item remove network partition
    \item verify max number of goroutines has not spiked wrt maxSeen
\end{enumerate*}

\begin{figure}[htbp]
    \centering
    \includegraphics[page=1, width=\columnwidth]{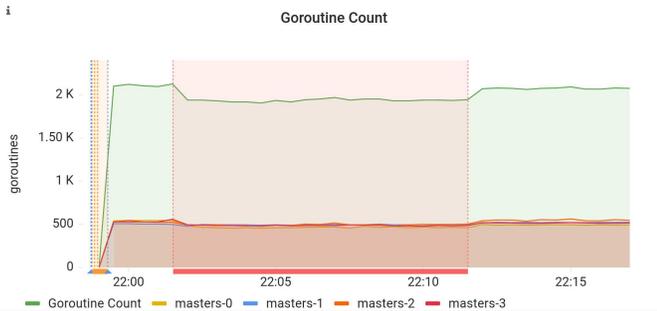}
    \caption{Network Partition}
    \label{fig:baseline:goroutines_leaking}
\end{figure}

As can be seen in Figure~\ref{fig:baseline:goroutines_leaking}, \frisbee{} can successfully inject and repair a network partition fault. However, the number of goroutines after the partition have not spiked, and \frisbee{} correctly continues the execution of the test.

\section{Related Work}
    \frisbee{} adheres to the two rules of end-to-end testing: help developers focus on the test case rather than test the testing mechanisms, and create similar environments for dev, test, and production. We compare the related work concerning these two rules. 
    
    \textbf{Test source code:}
    Frameworks like Ginkgo for Go, Robot for python, or Serenity for Java, provide a rich and expressive DSL (Domain-specific Language) for writing test scenarios. In return, the framework engine implements the scenario using thin threads, each representing a single service~\cite{bai2001distributed,tsai2001end}. However, this approach is impractical for systems whose components are written in different languages. 
    
    \noindent
    \textbf{Test processes in physical infrastructure:}
    Despite its simplicity, this approach comes with many known deficits, drifting library versions, shared namespaces on a host (e.g., two services cannot listen on the same port), library inconsistencies, and limited intervention on a shared infrastructure. Chef and Puppet provide declarative to automate the deployment and configuration of testing objects (e.g., servers, benchmarks). However, since these tools are designed for deployment and configuration, they expect the `final' state of the system. In contrast, testing generally involves multiple `intermediate' steps with logical dependencies between them. 
    

    \noindent
    \textbf{Test containers in Docker}
    With the popularisation of Docker, the testing community turned into replacing virtual machines with containers. In the early stages, containers were managed by scripts. This manual approach was cumbersome and mixed the test case with the testing mechanism. Later on, the Docker-native approach followed by BenchPilot~\cite{georgioubenchpilot}, Fogify~\cite{symeonides2020fogify}, and IOTier~\cite{nikolaidis2021iotier}, was to use the Docker-Compose language for running multi-container tests. However, besides lacking assertions in the language, Docker-compose works for single-node deployments. 
    
    
    \noindent
    \textbf{Test containers natively in Kubernetes}
    KUTTL~\cite{kuttl} and Iter8~\cite{10.1145/3472883.3486984} are Kubernetes-native testing tools, but their goals is different from ours. KUTTL is a toolkit for testing a Kubernetes controller, not a Kubernetes controller for testing other systems. Iter8 aims at automating the progressive rollout of microservices. To do so, they create two experiments. One Iter8 experiment for deploying microservices instrumented with telemetry agents, and one chaos experiment (via Litmus) for periodically killing the deployed microservices. However, since there is no coordination between the two experiments, the chaos experiment is execution-agnostic, and the final result is hardly reproducible. 

\section{Conclusion}
    Because our systems and system designs have evolved in such an unprecedented way, we must also evolve our testing methods to better understand how our systems perform under normal and adverse operating conditions.  We believe the answer lies in building a new generation testing tools that can support truly precise, controllable, observable, and language-agnostic experiments. With \frisbee{}, our ultimate goal is to foster the development of a common Cloud-Native Testing Framework~\cite{nikolaidis2021frisbee} for systems researchers and practitioners, in order to support dependable, reproducible, and comparable experiments in Chaos campaigns. The \frisbee{} and the experiments are made available open-source at https://frisbee.dev.

\section*{Acknowledgement}
    This project has received funding from the European Union's Horizon 2020 research and innovation programme under the Marie Sk\l{}odowska-Curie, grant agreement No. 894204 (Ether, H2020-MSCA-IF-2019).

\bibliographystyle{IEEEtran}
\bibliography{bibliography}

\begin{thebibliography}{10}
\providecommand{\url}[1]{#1}
\csname url@samestyle\endcsname
\providecommand{\newblock}{\relax}
\providecommand{\bibinfo}[2]{#2}
\providecommand{\BIBentrySTDinterwordspacing}{\spaceskip=0pt\relax}
\providecommand{\BIBentryALTinterwordstretchfactor}{4}
\providecommand{\BIBentryALTinterwordspacing}{\spaceskip=\fontdimen2\font plus
\BIBentryALTinterwordstretchfactor\fontdimen3\font minus
  \fontdimen4\font\relax}
\providecommand{\BIBforeignlanguage}[2]{{%
\expandafter\ifx\csname l@#1\endcsname\relax
\typeout{** WARNING: IEEEtran.bst: No hyphenation pattern has been}%
\typeout{** loaded for the language `#1'. Using the pattern for}%
\typeout{** the default language instead.}%
\else
\language=\csname l@#1\endcsname
\fi
#2}}
\providecommand{\BIBdecl}{\relax}
\BIBdecl

\bibitem{symeonides2020fogify}
M.~Symeonides, Z.~Georgiou, D.~Trihinas, G.~Pallis, and M.~D. Dikaiakos,
  ``Fogify: A fog computing emulation framework,'' in \emph{2020 IEEE/ACM
  Symposium on Edge Computing (SEC)}.\hskip 1em plus 0.5em minus 0.4em\relax
  IEEE, 2020, pp. 42--54.

\bibitem{nikolaidis2021iotier}
F.~Nikolaidis, M.~Marazakis, and A.~Bilas, ``Iotier: A virtual testbed to
  evaluate systems for iot environments,'' in \emph{2021 IEEE/ACM 21st
  International Symposium on Cluster, Cloud and Internet Computing
  (CCGrid)}.\hskip 1em plus 0.5em minus 0.4em\relax IEEE, 2021, pp. 676--683.

\bibitem{kao1996define}
W.-L. Kao and R.~K. Iyer, ``Define: A distributed fault injection and
  monitoring environment,'' in \emph{Proceedings of IEEE Workshop on
  Fault-Tolerant Parallel and Distributed Systems}.\hskip 1em plus 0.5em minus
  0.4em\relax IEEE, 1996, pp. 252--259.

\bibitem{abedi2017conducting}
A.~Abedi and T.~Brecht, ``Conducting repeatable experiments in highly variable
  cloud computing environments,'' in \emph{Proceedings of the 8th ACM/SPEC on
  International Conference on Performance Engineering}, 2017, pp. 287--292.

\bibitem{leitner2016patterns}
P.~Leitner and J.~Cito, ``Patterns in the chaos—a study of performance
  variation and predictability in public iaas clouds,'' \emph{ACM Transactions
  on Internet Technology (TOIT)}, vol.~16, no.~3, pp. 1--23, 2016.

\bibitem{giuffrida2013edfi}
C.~Giuffrida, A.~Kuijsten, and A.~S. Tanenbaum, ``Edfi: A dependable fault
  injection tool for dependability benchmarking experiments,'' in \emph{2013
  IEEE 19th Pacific Rim International Symposium on Dependable Computing}.\hskip
  1em plus 0.5em minus 0.4em\relax IEEE, 2013, pp. 31--40.

\bibitem{mongobug}
\BIBentryALTinterwordspacing
CockroachDB. (2022, Aug.) Recovery problems after network partition. [Online].
  Available: \url{https://jira.mongodb.org/browse/SERVER-23003}
\BIBentrySTDinterwordspacing

\bibitem{ng2001design}
W.~T. Ng and P.~M. Chen, ``The design and verification of the rio file cache,''
  \emph{IEEE Transactions on Computers}, vol.~50, no.~4, pp. 322--337, 2001.

\bibitem{simao2010fault}
A.~Sim{\~a}o and A.~Petrenko, ``Fault coverage-driven incremental test
  generation,'' \emph{The Computer Journal}, vol.~53, no.~9, pp. 1508--1522,
  2010.

\bibitem{10.1145/3447851.3458738}
\BIBentryALTinterwordspacing
F.~Nikolaidis, A.~Chazapis, M.~Marazakis, and A.~Bilas, ``Frisbee: A suite for
  benchmarking systems recovery,'' in \emph{Proceedings of the 1st Workshop on
  High Availability and Observability of Cloud Systems}, ser. HAOC '21.\hskip
  1em plus 0.5em minus 0.4em\relax New York, NY, USA: Association for Computing
  Machinery, 2021, p. 18–24. [Online]. Available:
  \url{https://doi.org/10.1145/3447851.3458738}
\BIBentrySTDinterwordspacing

\bibitem{jepsen}
\BIBentryALTinterwordspacing
Jepsen. (2020, Aug.) A framework for distributed systems verification, with
  fault injection. [Online]. Available:
  \url{https://github.com/jepsen-io/jepsen}
\BIBentrySTDinterwordspacing

\bibitem{10.1145/1807128.1807153}
\BIBentryALTinterwordspacing
G.~Candea, S.~Bucur, and C.~Zamfir, ``Automated software testing as a
  service,'' in \emph{Proceedings of the 1st ACM Symposium on Cloud Computing},
  ser. SoCC '10.\hskip 1em plus 0.5em minus 0.4em\relax New York, NY, USA:
  Association for Computing Machinery, 2010, p. 155–160. [Online]. Available:
  \url{https://doi.org/10.1145/1807128.1807153}
\BIBentrySTDinterwordspacing

\bibitem{10.1007/978-3-030-58858-8_26}
M.~Dole{\v{z}}el, ``Defining testops: Collaborative behaviors and
  technology-driven workflows seen as enablers of effective software testing in
  devops,'' in \emph{Agile Processes in Software Engineering and Extreme
  Programming -- Workshops}, M.~Paasivaara and P.~Kruchten, Eds.\hskip 1em plus
  0.5em minus 0.4em\relax Cham: Springer International Publishing, 2020, pp.
  253--261.

\bibitem{miller2021four}
S.~Miller and D.~Firesmith, ``Four types of shift left testing,''
  CARNEGIE-MELLON UNIV PITTSBURGH PA, Tech. Rep., 2021.

\bibitem{tolosana2010adaptive}
R.~Tolosana-Calasanz, J.~A. Banares, O.~F. Rana, P.~{\'A}lvarez, J.~Ezpeleta,
  and A.~Hoheisel, ``Adaptive exception handling for scientific workflows,''
  \emph{Concurrency and computation: Practice and experience}, vol.~22, no.~5,
  pp. 617--642, 2010.

\bibitem{kasuya2007verification}
A.~Kasuya and T.~Tesfaye, ``Verification methodologies in a tlm-to-rtl design
  flow,'' in \emph{2007 44th ACM/IEEE Design Automation Conference}.\hskip 1em
  plus 0.5em minus 0.4em\relax IEEE, 2007, pp. 199--204.

\bibitem{operator}
\BIBentryALTinterwordspacing
Kubernetes. (2021, Aug.) Operator best practices. [Online]. Available:
  \url{https://sdk.operatorframework.io/docs/best-practices/best-practices/}
\BIBentrySTDinterwordspacing

\bibitem{burns2018designing}
B.~Burns, \emph{Designing Distributed Systems: Patterns and Paradigms for
  Scalable, Reliable Services}.\hskip 1em plus 0.5em minus 0.4em\relax "
  O'Reilly Media, Inc.", 2018.

\bibitem{196346}
\BIBentryALTinterwordspacing
B.~Burns and D.~Oppenheimer, ``Design patterns for container-based distributed
  systems,'' in \emph{8th {USENIX} Workshop on Hot Topics in Cloud Computing
  (HotCloud 16)}.\hskip 1em plus 0.5em minus 0.4em\relax Denver, CO: {USENIX}
  Association, Jun. 2016. [Online]. Available:
  \url{https://www.usenix.org/conference/hotcloud16/workshop-program/presentation/burns}
\BIBentrySTDinterwordspacing

\bibitem{herbein2019mcem}
S.~Herbein, D.~Domyancic, P.~Minner, I.~Laguna, R.~da~Silva, and D.~Ahn,
  ``Mcem: Multi-level cooperative exception model,'' Lawrence Livermore
  National Lab.(LLNL), Livermore, CA (United States), Tech. Rep., 2019.

\bibitem{gupta2009cifts}
R.~Gupta, P.~Beckman, B.-H. Park, E.~Lusk, P.~Hargrove, A.~Geist, D.~Panda,
  A.~Lumsdaine, and J.~Dongarra, ``Cifts: A coordinated infrastructure for
  fault-tolerant systems,'' in \emph{2009 International Conference on Parallel
  Processing}.\hskip 1em plus 0.5em minus 0.4em\relax IEEE, 2009, pp. 237--245.

\bibitem{di2017logaider}
S.~Di, R.~Gupta, M.~Snir, E.~Pershey, and F.~Cappello, ``Logaider: A tool for
  mining potential correlations of hpc log events,'' in \emph{2017 17th
  IEEE/ACM International Symposium on Cluster, Cloud and Grid Computing
  (CCGRID)}.\hskip 1em plus 0.5em minus 0.4em\relax IEEE, 2017, pp. 442--451.

\bibitem{movatic}
\BIBentryALTinterwordspacing
CockroachDB. (2022, Mar.) Movatic deploys cockroachdb in kubernetes for iot
  workloads. [Online]. Available:
  \url{https://www.cockroachlabs.com/customers/movatic/}
\BIBentrySTDinterwordspacing

\bibitem{chaomesh}
\BIBentryALTinterwordspacing
PingCAP. (2020) A chaos engineering platform for kubernetes. [Online].
  Available: \url{https://github.com/chaos-mesh/chaos-mesh}
\BIBentrySTDinterwordspacing

\bibitem{roach-test-ssstable-corruption}
\BIBentryALTinterwordspacing
CockroachDB. (2022, Mar.) Sstable corruption. [Online]. Available:
  \url{https://github.com/cockroachdb/.../roachtest/tests/sstable\_corruption.go}
\BIBentrySTDinterwordspacing

\bibitem{roach-test-goroutines-partition}
\BIBentryALTinterwordspacing
------. (2022, Mar.) Sstable corruption. [Online]. Available:
  \url{https://github.com/cockroachdb/.../roachtest/tests/network.go}
\BIBentrySTDinterwordspacing

\bibitem{bai2001distributed}
X.~Bai, W.-T. Tsai, R.~Paul, T.~Shen, and B.~Li, ``Distributed end-to-end
  testing management,'' in \emph{Proceedings Fifth IEEE International
  Enterprise Distributed Object Computing Conference}.\hskip 1em plus 0.5em
  minus 0.4em\relax IEEE, 2001, pp. 140--151.

\bibitem{tsai2001end}
W.-T. Tsai, X.~Bai, R.~Paul, W.~Shao, and V.~Agarwal, ``End-to-end integration
  testing design,'' in \emph{25th Annual International Computer Software and
  Applications Conference. COMPSAC 2001}.\hskip 1em plus 0.5em minus
  0.4em\relax IEEE, 2001, pp. 166--171.

\bibitem{georgioubenchpilot}
J.~Georgiou, M.~Symeonides, M.~Kasioulis, D.~Trihinas, G.~Pallis, and M.~D.
  Dikaiakos, ``Benchpilot: Repeatable \& reproducible benchmarking for edge
  micro-dcs.''

\bibitem{kuttl}
\BIBentryALTinterwordspacing
KUTTL. (2022, Mar.) What is kuttl? [Online]. Available: \url{The KUbernetes
  Test TooL}
\BIBentrySTDinterwordspacing

\bibitem{10.1145/3472883.3486984}
\BIBentryALTinterwordspacing
M.~Toslali, S.~Parthasarathy, F.~Oliveira, H.~Huang, and A.~K. Coskun,
  \emph{Iter8: Online Experimentation in the Cloud}.\hskip 1em plus 0.5em minus
  0.4em\relax New York, NY, USA: Association for Computing Machinery, 2021, p.
  289–304. [Online]. Available: \url{https://doi.org/10.1145/3472883.3486984}
\BIBentrySTDinterwordspacing

\bibitem{nikolaidis2021frisbee}
F.~Nikolaidis, A.~Chazapis, M.~Marazakis, and A.~Bilas, ``Frisbee: automated
  testing of cloud-native applications in kubernetes,'' \emph{arXiv preprint
  arXiv:2109.10727}, 2021.

\end{thebibliography}

\end{document}